\documentstyle [12pt]{article}

\newcommand{\lbl}[1]{\label{eq: #1}}

\def\R{{\rm I\hspace{-.15em}R}}

\def\C{\hspace{3pt}{\rm l\hspace{-.47em}C}}

\def\1{\mathbf {Id} }
\def\z {{\cal Z}}

\def\dz {\frac{d}{d{\cal Z}}\,}

\def\ab{\frac{\partial x^{\alpha}}{\partial X^a}\frac{\partial x'^{\beta'}}{\partial X'^{b'}}\;}

\def\z {{\cal Z}}

\def\b{\begin{equation}}
\def\e{\end{equation}}
\def\bd{\begin{displaystyle}}
\def\ed{\end{displaystyle}}
\def\ba{\begin{array}}
\def\ea{\end{array}}

\def\bee{\begin{enumerate}}
\def\eee{\end{enumerate}}

\def\bes{\begin{eqnarray*}}
\def\ees{\end{eqnarray*}}
\def\be{\begin{eqnarray}}
\def\ee{\end{eqnarray}}

\begin{document}

\title{Conformally invariant wave-equations\\ and massless fields in de Sitter spacetime}

\author{S. Behroozi$^1$, S. Rouhani$^2$, M.V. Takook$^{1}$\thanks{e-mail:
takook@razi.ac.ir}\\ and M.R. Tanhayi$^2$}

\maketitle   \centerline{\it $^1$Department of Physics, Razi
University, Kermanshah, IRAN} \centerline{\it $^2$Plasma Physics
Research Center, Islamic Azad University,}
 \centerline{\it P.O.BOX 14835-157, Tehran,
IRAN}
 \vspace{15pt}

\begin{abstract}

Conformally invariant wave equations in de Sitter space, for
scalar and vector fields, are introduced in the present paper.
Solutions of their wave equations and the related two-point
functions, in the ambient space notation, have been calculated.
The ``Hilbert'' space structure and the field operator, in terms
of coordinate independent de Sitter plane waves, have been
defined. The construction of the paper is based on the analyticity
in the complexified pseudo-Riemanian manifold, presented first by
Bros et al.. Minkowskian limits of these functions are analyzed.
The relation between the ambient space notation and the intrinsic
coordinates is then studied in the final stage.

\end{abstract}

\vspace{0.5cm} {\it Proposed PACS numbers}: 04.62.+v, 11.25.Hf,
11.10.Cd, 98.80.Hw \vspace{0.5cm}

\newpage
\section{Introduction}

Recent astrophysical data indicate that our universe might currently
be in a de Sitter (dS) phase. Quantum field theory in dS space-time
has evolved as an exceedingly important subject, studied by many
authors in the course of the past decade. The importance of dS space
has been primarily ignited by the study of the inflationary model of
the universe and the quantum gravity. The importance of the
''massless'' spin-$2$ field in the de Sitter space (dS linear
quantum gravity) is due to the fact that it plays the central role
in quantum gravity and quantum cosmology. Massless field equations
in de Sitter space, similar to the flat space counterparts, have the
conformal invariance properties. The massless field equations with
$s\geq1$ are also gauge invariant. In this paper, the conformally
invariant aspects of the massless scalar field and the massless
spin-$1$ field (vector field) in dS space are studied. This
formulation establishes the base for conformally invariant wave
equation of massless spin-$2$ field.

Bros et al. \cite{brgamo,brmo} presented QFT of scalar field in dS
space that closely mimics QFT in Minkowski space. They have
introduced a new version of the Fourier-Bros transformation on the
hyperboloid \cite{brmo2}, which allows us to completely
characterize the Hilbert space of ``one-particle'' state and the
corresponding irreducible unitary representations of the de Sitter
group. In this construction, correlation functions are boundary
values of analytical functions. It should be noted that the
analyticity condition is only preserved in the case of Euclidean
vacuum. In a series of papers we generalized the Bros construction
to the quantization of the various spin free fields in dS space
\cite{ta}. Here we have applied the Bros construction to the
conformally invariant massless scalar  and vector fields in dS
space.

The massive and massless conformally coupled scalar fields,
respectively correspond to the principal and complementary series
representation of de Sitter group \cite{brmo}. The massive and
massless vector fields in dS space have been associated with the
principal series and the lowest representation in the vector
discrete series representation of dS group, respectively
\cite{gata,garota}. These representations have the physical meaning
in the null curvature limit.  The massless vector field, however,
with the divergencelessness condition, is singular \cite{gata}. This
type of singularity is actually due to the divergencelessness
condition needed to associate this field with a specific unitary
irreducible representation (UIR) of dS group. To solve this problem,
the divergencelessness condition must be dropped. The field equation
is then gauge invariant \cite{garota}. Hence the vector field is
associated with an indecomposable representation of the dS group. By
fixing the gauge, this field can be quantized. In this case,
emergence of states with negative or null norms necessitates
indefinite metric quantization \cite{garota}. In order to eliminate
these unphysical states, certain conditions must be imposed on the
field operators and on the vacuum state, similar to the pattern of
Minkowskian space theories \cite{wiga}. Physical states propagate on
the light-cone and correspond to the vector massless Poincar\'e
field in the null curvature limit. It has been proven that the use
of an indefinite metric is unavoidable if one insists on the
preservation of causality (locality) and covariance in gauge quantum
field theories \cite{st}. The generalization of the Wightman axioms
to the QFT in de Sitter space, for scalar, spinor and vector field,
has been studied by Bros, Gazeau et al \cite{brmo,ta,gata,garota}.

The free massless de Sitter vector field in the flat coordinate
system has been studied previously \cite{bodu}. This covers only
one-half of the dS hyperboloid. In $1986$ Allen calculated the
massless vector two-point functions in terms of a maximally
symmetric bi-tensor. His simple choice of gauge broke the
conformal invariance and led to the appearance of logarithmic
singularity \cite{alja}.

In the section $2$, we have briefly recalled the main result of the
previous paper \cite{garota}, {\it i.e.} the gauge invariant dS
vector field equation in terms of the Casimir operator. The six-cone
formalism is presented in section $3$. Following description of de
Sitter coordinates, the projection techniques have been introduced.
Conformally invariant wave equations have been obtained in the next
stage. Section $4$ is devoted to the solutions of the field
equations in terms of a de Sitter plane wave and a polarization
vector ${\cal E}_\alpha$. Due to the presence of a multivalued phase
factor and the presence of a singularity, these solutions are not
globally defined. Extending these solutions to the complex dS space,
have allowed us to circumvent these problems altogether \cite{brmo}.
The two-point functions are calculated in section $5$. The
``Hilbert'' space structure and the field operators, in terms of
coordinate independent dS plane waves, have been defined in this
section. The null curvature limit of the two-point functions and the
relation between the ambient space notation and the intrinsic
coordinates are studied in the next stage. Finally, a brief
conclusion and an outlook have been given in section $6$.

\setcounter{equation}{0}
\section{de Sitter field equations}

The de Sitter space-time can be defined by the one-sheeted
four-dimensional hyperboloid:
     \b X_H=\{x \in \R^5:\;x^2=\eta_{\alpha\beta} x^\alpha x^\beta =-H^{-2}\},\;\;
      \alpha,\beta=0,1,2,3,4, \e
where $\eta_{\alpha\beta}=$diag$(1,-1,-1,-1,-1)$. The de Sitter
metric is
  \b  ds^2=\eta_{\alpha\beta}dx^{\alpha}dx^{\beta}\mid_{x^2 =-H^{-2}}=
     g_{\mu\nu}^{dS}dX^{\mu}dX^{\nu},\;\; \nu,\mu=0,1,2,3,\e
where $X^\mu$ are the $4$ space-time coordinates in dS hyperboloid
and $x^{\alpha}$ are the $5$-dimensional coordinates in the
ambient space notation. For simplicity one can put $H=1$. The wave
equation for massless conformally coupled scalar field is \b
(\Box_H+2)\phi=0,\e where $\Box_H$ is the Laplace-Beltrami
operator on dS space. In the ambient space notation, the wave
equation is written in the following form \cite{brmo} \b (
Q^{(0)}-2)\phi=0,\e where $Q^{(0)}$ is the second order scalar
Casimir operator of de Sitter group $SO_{0}(1,4)$. The covariant
derivative of a tensor field, $T_{\alpha_1....\alpha_n}$, in the
ambient space notation is \b \nabla_\beta
T_{\alpha_1....\alpha_n}=\bar
\partial_\beta
T_{\alpha_1....\alpha_n}-\sum_{i=1}^{n}x_{\alpha_i}T_{\alpha_1..\alpha_{i-1}\beta\alpha_{i+1}..\alpha_n},\e
where $\bar
\partial_\alpha=\theta_{\alpha \beta}\partial^\beta=
\partial_\alpha+x_\alpha(x. \partial)$ and $\theta_{\alpha
\beta}$ is the transverse projector ($\theta_{\alpha \beta}=
\eta_{\alpha \beta}+ x_\alpha x_\beta$). In terms of the covariant
derivative, the second order scalar Casimir operator is
$Q^{(0)}=-\bar
\partial^2$. The wave equation for massless vector fields $A_\mu(X)$
propagating on dS space gives \cite{alja} \b (\Box_H +3)
A_\mu(X)-\nabla_\mu \nabla . A=0.\e This field equation is
invariant under the gauge transformation $ A_{\mu} \longrightarrow
A_{\mu}^{gt}=A_{\mu}+\nabla_{\mu}\phi_g,$ where $\phi_g$ is an
arbitrary scalar field. The gauge-fixed wave equation is
\cite{alja} \b (\Box_H +3) A_\mu(X)-c\nabla_\mu \nabla . A=0,\e
where $c$ is a gauge-fixing parameter. It is an arbitrary positive
real number.

In order to simplify the relation between the field and the
representation of the dS group, we have adopted the vector field
notation $K_{\alpha}(x)$ in ambient space notation. Pursuing this
notation, the solutions of the field equations are easily written
in terms of scalar fields. Consequently a gauge transformation has
vividly appeared. The $4$-vector field $A_{\mu}(X)$ is locally
determined by the $5$-vector field $K_{\alpha}(x)$ through the
relation \b\lbl{passage} A_{\mu}(X)=\frac{\partial
x^{\alpha}}{\partial X^{\mu}}K_{\alpha}(x(X)). \e Using the
equation $(2.8)$ and the transversality condition ($x\cdot K=0$),
we have $$ \nabla_\rho \nabla_\mu A_\nu=\frac{\partial
x^{\gamma}}{\partial X^{\rho}}\frac{\partial x^{\alpha}}{\partial
X^{\mu}}\frac{\partial x^{\beta}}{\partial X^{\nu}}[ \bar
\partial_\gamma(\bar \partial_\alpha K_{\beta}-x_\beta
K_\alpha)$$ \b -x_\alpha (\bar \partial_\gamma K_{\beta}-x_\beta
K_\gamma)-x_\beta (\bar \partial_\alpha K_\gamma-x_\gamma
K_\alpha)].\e Using above equations, the field equation $(2.6)$,
in the ambient space notation, gives \cite{gata,garota}: \b ((\bar
\partial)^2+2)K(x)-2x\bar
\partial.K(x)-\bar \partial \partial.K=0.\e In terms of
the second order vector Casimir operator $Q^{(1)}$, one obtains
\cite{gata,garota}: \b Q^{(1)}K(x)+D_1 \partial.K=0,\:\:\:x.K=0
,\e where $D_1=\bar
\partial$. The Casimir
operator $Q_1^{(1)}$ is defined by $$   Q_1^{(1)}=-\frac{1}{2}
L^{\alpha\beta}L_{\alpha\beta}=-\frac{1}{2}
(M^{\alpha\beta}+S^{\alpha\beta})(M_{\alpha\beta}+S_{\alpha\beta}),$$
where $M_{\alpha\beta}=-i (x_\alpha \partial_\beta-x_\beta
\partial_\alpha)=-i (x_\alpha \bar \partial_\beta-x_\beta \bar
\partial_\alpha)$ and the action of the spin generator
$S_{\alpha\beta}$ is defined by $$
S_{\alpha\beta}K_\gamma=-i(\eta_{\alpha\gamma}K_{\beta}-\eta_{\beta\gamma}K_\alpha).$$

The field equation is gauge invariant, {\it i.e.}
   \b K\longrightarrow K^{gt}=K+D_1\phi_g \Longrightarrow
Q^{(1)}K^{gt}(x)+D_1\partial.K^{gt}=0, \e where $\phi_g$ is an
arbitrary scalar field. The gauge-fixed wave equation in the
ambient space notation is \b Q^{(1)}K(x)+cD_1
\partial.K=0. \e This could be directly obtained from $(2.7$).

If we consider the physical subspace of solutions with $\partial.
K=0$, we have \b Q^{(1)}K(x)=0=(Q^{(0)}-2)K(x),\e which is similar
to the equation $(2.4)$. This field can be associated with the UIR's
$\Pi^\pm_{1,1}$ of the dS group. For obtaining the solution of the
equation $(2.14)$ following of the massive field case, the principal
series parameter ($\nu $) must be replaced by $\pm \frac{i}{2}$
\cite{garota}.  Replacement of principal series parameter by
discrete series results in appearance of singularities in vector
field solution. This singularity is actually due to the
divergencelessness condition needed to associate this field with a
specific UIR of the dS group \cite{garota}. In section $4$, we
calculate the general solutions of the field equations for the
different values of $c$. In the next section we have obtained the
specific value of $c$, that makes the wave equation, conformally
invariant.

\setcounter{equation}{0}
\section{Conformally invariant wave equations}

In the Minkowski space, the massless field equations are conformally
invariant. For every massless representation of Poincar\'e group
there exists only one corresponding representation in the conformal
group \cite{babo,anla}. In the de Sitter space, for the vector
field, only two representations in the discrete series
$(\Pi^{\pm}_{1,1})$ have a Minkowskian interpretation. The signs
$\pm$ correspond to two types of helicity for the massless vector
field. The representation $\Pi^+_{1,1}$ has a unique extension to a
direct sum of two UIR's ${\cal C}(2;1,0)$ and ${\cal C}(-2;1,0)$ of
the conformal group $SO_0(2,4)$. Note that  ${\cal C}(2;1,0)$ and
${\cal C}(-2;1,0)$ correspond to positive and negative energies
representation in the conformal group respectively \cite{babo,anla}.
The concept of energy cannot be defined in de Sitter space. The
latter restricts to the vector massless Poincar\'e UIR's $P^>(0, 1)$
and $P^<(0,1)$ with positive and negative energies respectively. The
following diagrams illustrate these connections
$$ \left. \begin{array}{ccccccc}
     &             & {\cal C}(2,1,0)
& &{\cal C}(2,1,0)   &\hookleftarrow &{\cal P}^{>}(0,1)\\
 \Pi^+_{1,1} &\hookrightarrow  & \oplus
&\stackrel{H=0}{\longrightarrow} & \oplus  & &\oplus  \\
     &             & {\cal C}(-2,1,0)&
& {\cal C}(-2,1,0)  &\hookleftarrow &{\cal P}^{<}(0,1),\\
    \end{array} \right. $$
$$ \left. \begin{array}{ccccccc}
     &             & {\cal C}(2,0,1)
& &{\cal C}(2,0,1) &\hookleftarrow &{\cal P}^{>}(0,-1)\\
 \Pi^-_{1,1} &\hookrightarrow  & \oplus
&\stackrel{H=0}{\longrightarrow} &  \oplus & &\oplus  \\
     &             & {\cal C}(-2,0,1)&
& {\cal C}(-2,0,1)   &\hookleftarrow &{\cal       P}^{<}(0,-1),\\
    \end{array} \right. $$
where the arrows $\hookrightarrow $ designate unique extension. $
{\cal P}^{ \stackrel{>}{<}}(0,-1)$ are the massless Poincar\'e UIR
with positive and negative energies and negative helicity. In this
section, the conformal invariance of massless scalar and vector
fields in de Sitter space is studied. Conformally invariant wave
equations are best obtained by the use of Dirac's null-cone in
$\R^6$, followed by the projection of the equations to the de
Sitter space \cite{dir}.

\subsection{Dirac's six-cone formalism}

Dirac's six-cone is a $5$-dimensional super-surface \b
u^2=\eta_{ab}u^au^b=0, \;\;
\eta_{ab}=\mbox{diag}(1,-1,-1,-1,-1,1),\e in $\R^6$, where
$a,b=0,1,2,3,4,5$. An operator $\hat{A}$ which acts on the field
$\phi$, over $\R^6$, is said to be intrinsic if
\cite{gaha,bifrhe,gahamu} \b \hat{A}u^2 \phi=u^2 \hat{A'}\phi, \;\;
\mbox{for any $\phi$}.\e The following are examples of the intrinsic
operators:
\begin{enumerate}

\item{ Fifteen generators of the conformal group $SO_0(2,4)$,}
$$M_{ab}=i(u_a\partial_b-u_b\partial_a).$$

\item{ The conformal-degree operator $N_5$} $$ N_5\equiv
u^a\partial_a.$$

\item{ The intrinsic gradient} $$Grad_a\equiv
u_a\partial_b\partial^b-(2N_5+4)\partial_a.$$

\item{The powers of d'Alembertian} $$(\partial_a\partial^a)^p,$$
which acts intrinsically on fields of conformal degree $(p-2)$.
\end{enumerate}

The following conformally invariant set of equations, on the cone,
has been used most commonly \b \left\{ \ba{rcl}
(\partial_a\partial^a)^p\Psi&=&0,\\
N_5\Psi&=&(p-2)\Psi,\ea\right.\e where $\Psi$ is a tensor field of
a definite rank and of a definite symmetry. The other conformally
invariant conditions can be added to the above system in order to
restrict the space of the solutions. The following conditions are
introduced to achieve the above goal:
\begin{enumerate}

\item{transversality} $$u_a\Psi^{ab...}=0 ,$$

\item{divergencelessness} $$Grad_a\Psi^{ab...}=0 ,$$

\item{tracelessness} $$\Psi_{ab...}^a=0 .$$
\end{enumerate}

\subsection{Projective of the six-cone}

In order to project the coordinates on the cone $u^2=0$, to the
$4+1$ de Sitter space we chose the following relation:\b \left\{
\ba{rcl}
x^{\alpha}&=&(u^5)^{-1}u^\alpha,\\
x^5&=&u^5.\ea\right.\e Note that $x^5$ becomes superfluous when we
deal with the projective cone. Various intrinsic operators
introduced in previous section now read as:

\begin{enumerate}

\item{ the ten $SO_0(1,4)$ generators} \b
M_{\alpha\beta}=i(x_{\alpha}\partial_{\beta}-x_{\beta}\partial_{\alpha}),
\e

\item{the conformal-degree operator $(N_5)$} \b N_5=
 x_5\frac{\partial}{\partial x_5},\e

\item{the conformal gradient $(Grad_{\alpha})$} \cite{gahamu} \b
Grad_{\alpha}= -x_{5}^{-1}
\{x_{\alpha}[Q^{(0)}-N_{5}(N_{5}-1)]+2\bar{\partial}_{\alpha}(N_{5}+1)\},\e

\item{the powers of d'Alembertian $(\partial_a\partial^a)^p$},
which acts intrinsically on field of conformal degree $(p-2)$, \b
(\partial_{a}
\partial^{a})^{p}=-x_{5}^{-2p} \prod_{j=1}^{p}[Q^{(0)}+(j+1)(j-2)], \e
\end{enumerate}

\subsection{Conformally invariant equations}

For scalar field, the simplest conformally invariant system is
obtained from $(3.3)$ with $p=1$,\b \left\{ \ba{rcl}
(\partial_a\partial^a)\Psi&=&0,\\
N_5\Psi&=&-\Psi,\ea\right.\e where $\Psi$ is a scalar field on the
cone. We introduce the scalar de Sitter field by $\phi=x_{5}\Psi$.
This obeys the conformally invariant equation derived from $(3.8)$
and $(3.9)$: \b (Q^{(0)}-2)\phi=0,\e which is a massless
conformally coupled scalar field in de Sitter space (Eq. $(2.4)$).

Similarly, the conformally invariant system for vector field is
obtained from $(3.3)$ with $p=1$. In this case $\Psi$ is a tensor
of rank one, and it is assumed to be the solution of \b \left\{
\ba{rcl}
(\partial_a\partial^a)\Psi_a&=&0,\\
N_5\Psi_a&=&-\Psi_a.\ea\right.\e We classify the six degrees of
freedom of the vector fields on the cone by, \b
K_{\alpha}=x_{5}(\Psi_{\alpha}+x_{\alpha}x\cdot\Psi),\;\;
\phi_1=x_{5}\Psi_{5}, \;\; \phi_2=x_{5} x\cdot\Psi,\e where
$K_{\alpha}$ is a vector field on de Sitter space ($x.K=0$). Using
the equations $(3.8)$ and $(3.11)$, these fields are proved to obey
the following conformal system of equations: \b
Q^{(1)}K_{\alpha}+\frac{2}{3}D_{1\alpha}\bar{\partial}.K+\frac{1}{6}D_{1\alpha}
Q^{(0)}\bar{\partial}.K=0, \e \b (Q^{(0)}-2)\phi_1=0, \e \b
\phi_2=\frac{x_5}{12}(Q^{(0)}+4)\bar{\partial}.K. \e The divergence
of $(3.13)$ leads to \b Q^{(0)}(Q^{(0)}-2)\bar{\partial}.K=0.\e
Adding the conformal invariance condition \b
u^a\Psi_a=x^5(x.\Psi+\Psi_5)=0,\e to the above relations, we obtain
the following conformal systems: \b
Q^{(1)}K_{\alpha}+D_{1\alpha}\bar{\partial}.K=0, \e \b
(Q^{(0)}-2){\partial}.K=0, \e which correspond to the gauge fixing
$c=1$ in $(2.13)$. This is not a fully gauge invariant case, since
condition $(3.19)$ restricts the gauge field space. Nonetheless it
preserves the null-cone propagation of the solutions. Under the
gauge transformation  $K^{gt}=K+D_1\phi_g$, the equation $(3.19)$
requires the scalar field $\phi_g$ to satisfies the following
equation (this is indeed not an arbitrary scalar field) \b
Q^{(0)}(Q^{(0)}-2)\phi_g=0.\e The general solutions of the field
equations are calculated in the next section.

\setcounter{equation}{0}
\section{Conformally invariant solutions}

A general solution of equation $(2.13)$ can be written in terms of
two scalar fields $\phi_1$ and $\phi_2$ \b  K_{\alpha}=\bar
Z_{\alpha}\phi_1+D_{1\alpha}\phi_2,\e where $Z$ is a constant
five-vector and $\bar Z_{\alpha}=\theta_{\alpha\beta} Z^{\beta}$.
The vector field solution was obtained in terms of a ''massless''
conformally coupled scalar field $\phi$ \cite{garota}, \b
K^c=\left(\bar Z -\frac{c}{2(1-c)}D_1[x.Z +Z.\bar \partial
]+\frac{2-3c}{1-c}D_1[Q^{(0)}]^{-1}x.Z \right)\phi,\; c\neq 1.\e If
we use the scalar dS plane wave for massless conformally coupled
scalar field \cite{brmo}, {\it i.e.}
$$ \phi=(x.\xi)^\sigma,\;\; \sigma=-1,-2 ,$$ dS vector-plane wave $(4.2)$ could
not be properly defined since its last term is singular,
$$[Q^{(0)}]^{-1}x.Z (x.\xi)^\sigma=\frac{-1}{(\sigma+1)(\sigma+4)}x.Z
(x.\xi)^\sigma, \;\; \sigma=-1.$$ The five-vector $\xi $ lies on
the null cone $ {\cal C} = \{ \xi \in \R^5;\;\; \xi^2=0 \}$. For
$c=\frac{2}{3}$, however, it can be properly defined
\cite{garota}: \b K_{\alpha}(x)= \left[\bar
Z_{\alpha}+(\sigma-1)\frac{Z.\xi} {( x.\xi)^2}\bar
\xi_{\alpha}+(\sigma+1)\frac{Z.x}{x.\xi} \bar
\xi_{\alpha}\right](x.\xi)^\sigma,\e  where $\bar
\xi_\alpha=\theta_{\alpha \beta}\xi^\beta$. It is clear that this
method could not be used for the conformally invariant case {\it
i.e.} $c=1$.

An alternative pattern is presented here for other values of $c$
corresponding to conformally invariant case ($c=1$), and physical
state ($\partial.K=0$ or $c=0$). The general solution of the field
equation can be written in terms of a generalized polarization
five-vector and a dS plane wave $$ K_{\alpha}(x)= {\cal
E}_{\alpha} (x,\xi,Z,\sigma)(x.\xi)^\sigma.$$ Using the pattern of
equation ($4.3)$, we introduce two constant parameters $b_1$ and
$b_2$ to define the following polarization vector
$${\cal E}_{\alpha}=\left[\bar Z_{\alpha}+b_1 \frac{Z.\xi} {(
x.\xi)^2}\bar \xi_{\alpha}+b_2\frac{Z.x}{x.\xi} \bar
\xi_{\alpha}\right]. $$ This vector satisfies the condition
$x.{\cal E}=0$. Imposing the condition $\partial . K=0$ to meet
the criterion of a physical state, {\it i.e.} divergenceless
condition, it is shown that the two constant parameters are
$$ b_1=\frac{1-\sigma}{2+\sigma},\;\;\;b_2=-1.$$ In this case the divergence of
polarization five-vector is $$ \partial.{\cal E}=-\sigma
\frac{\bar \xi.{\cal E}}{x.\xi}=
\frac{-3\sigma}{2+\sigma}\frac{Z.\xi}{x.\xi}.$$ The choice of
$Z.\xi=0$, which results in $ \partial.{\cal E}=0$, is a suitable
restriction on the arbitrary five-vector $Z^\alpha$, since it
renders a simplified solution and in the null curvature limit it
embarks on Minkowskian solution of the two-point function. By the
use of this condition ($Z.\xi=0$), the degrees of freedom for the
arbitrary five-vector field $Z^\alpha$ reduce to $4$. The
generalized polarization vector, then becomes
$${\cal E}_{\alpha}^d=\left[\bar Z_{\alpha}-\frac{Z.x}{x.\xi} \bar
\xi_{\alpha}\right]=\left[ Z_{\alpha}-\frac{Z.x}{x.\xi}
\xi_{\alpha}\right]. $$ This polarization vector satisfies the
interesting relation:
$$ Q^{(0)} {\cal E}_{\alpha}^d\phi={\cal E}_{\alpha}^dQ^{(0)}\phi.$$ If the vector
field $K$ does not satisfy the divergenceless condition, by the
choice of $Z.\xi=0$, it takes the following form $$ K_{\alpha}(x)=
{\cal E}_{\alpha} (x,\xi,Z,\sigma)(x.\xi)^\sigma ,\;\;\; {\cal
E}_{\alpha}=\left[\bar Z_{\alpha}+a\frac{Z.x}{x.\xi} \bar
\xi_{\alpha}\right]={\cal E}_{\alpha}^d+(1+a)\frac{Z.x}{x.\xi}
\bar \xi_{\alpha}, $$ where $a$ is an arbitrary constant
parameter. This parameter depends on the gauge parameter $c$ and
homogenous degree $\sigma$. If $K$ satisfies the vector field
equation $(2.13)$, the arbitrary constant parameter $a$ is fixed
at once. The divergence of the vector field, through the condition
$Z.\xi=0$, is
$$
\partial.K(x)=(1+a) (\sigma+4)x.Z (x.\xi)^\sigma.$$ This equation is itself a scalar
field with the homogenous degree $(\sigma+1)$ $$ Q^{(0)}
\partial.K(x)=-(\sigma+1)(\sigma+4)\partial.K(x).$$
Implementation of $K$ in the wave equation $(2.13)$ together with
the identity:
$$ Q^{(0)} x.Z (x.\xi)^\sigma x^\alpha=-(\sigma+2)(\sigma+5) x.Z (x.\xi)^\sigma x^\alpha-2
\left[ Z^{\alpha}+\sigma\frac{Z.x}{x.\xi}
\xi^{\alpha}\right](x.\xi)^\sigma,$$ results in the following
system of equations
\b\left\{%
\begin{array}{ll}
   a[ \sigma(\sigma+3)+2]-(1+a)\sigma(-2+c\sigma+4c)=0,& \hbox{(I)} \\
     \sigma(\sigma+3)+2-(1+a)(-2+c\sigma+4c)=0,& \hbox{(II)} \\
   (1+a)(\sigma+1)(\sigma+4)(1-c)=0.& \hbox{(III)} \\
\end{array}%
\right.\e All other values of $a$ and $\sigma$ can be categorized
in terms of various choices of the gauge parameter $c$. In the
present chapter three values of $c$ $(c=0,1, \frac{2}{3}$) have
been studied.

For $c=\frac{2}{3}$, the solutions of the system of the equations $(4.4)$ are $$ \left\{%
\begin{array}{ll}
    \sigma=-1, & \hbox{a=\mbox{arbitrary},} \\
    \sigma=-2, & \hbox{a=-1.} \\
\end{array}%
\right.$$ For the case $\sigma=-1$, value $a=0$ leads to the
previous solution Eq.($4.3)$. In this gauge, the solution of the
wave equation becomes \b K_\alpha=\left( \bar Z_\alpha+(\sigma+1)
\frac{Z.x}{x.\xi}\bar \xi_\alpha\right)( x.\xi)^\sigma=\left( \bar
Z_\alpha+\frac{\sigma+1}{\sigma} Z.x \bar
\partial_\alpha\right)(x.\xi)^\sigma.\e

For  $c=0$, the two solutions of the system of the equations
$(4.4)$
are $$ \left\{%
\begin{array}{ll}
    \sigma=-1, & \hbox{a=-1,} \\
    \sigma=-2, & \hbox{a=-1.} \\
\end{array}%
\right.$$ In this gauge the solution of the wave equation becomes:
\b K_\alpha=\left( \bar Z_\alpha-\frac{Z.x}{x.\xi}\bar
\xi_\alpha\right)(x.\xi)^\sigma=\left( \bar
Z_\alpha-\frac{1}{\sigma} Z.x \bar
\partial_\alpha\right)( x.\xi)^\sigma,\e which is clearly divergenceless. This field can be associated
to the UIR's $\Pi^\pm_{1,1}$ of the dS group and corresponds to
the physical state.

For $c=1$, the solutions of the system of the equations $(4.4)$ are \b \left\{%
\begin{array}{ll}
    \sigma=0, & \hbox{a=0,} \\
    \sigma=-1, & \hbox{a=-1,} \\
    \sigma=-2, & \hbox{a=\mbox{arbitrary},} \\
    \sigma=-3, & \hbox{a=-3.} \\
\end{array}%
\right. \e In this gauge, fixing $a$ to be $-2$ while $\sigma=-2$,
the solution results in $$ K_\alpha=\left( \bar Z_\alpha+\sigma
\frac{Z.x}{x.\xi}\bar \xi_\alpha\right)( x.\xi)^\sigma=\left( \bar
Z_\alpha+Z.x \bar
\partial_\alpha\right)( x.\xi)^\sigma.$$ The Eq.$(3.19)$ restricts the solutions
to the values $\sigma=-2,-3$, which are the conformally invariant
solutions.

It is more suitable to represent entire solutions of the field
equation in the following form: \b K_\alpha=\left( \bar
Z_\alpha+a(c,\sigma)\frac{Z.x}{x.\xi}\bar \xi_\alpha\right)(
x.\xi)^\sigma=\left( \bar
Z_\alpha+\frac{a(c,\sigma)}{\sigma}Z.x\bar
\partial_\alpha\right)( x.\xi)^\sigma.\e In contrast to
Minkowskian case, the generalized polarization vector ${\cal
E}_{\alpha}(x,\xi,Z,c,\sigma)$ is a function of the space-time $x$.
These solutions, however, are problematic as well. In contrast to
the ``massive'' field case in de Sitter space, the two solutions are
not complex conjugate of each other! We shall return to this point
when we construct the quantum field in the forthcoming chapter.
There also appears an arbitrary constant five-vector $Z$ (with one
constraint $Z.\xi=0$) in the solution of the field equation. This is
reminiscent of the problem of the vacuum state in the curved space.
For simplicity, we impose the condition that the solution in the
limit $H=0$ must be exactly the Minkowskian solution. This condition
in the massive scalar, spinor and vector cases, results in the
choice of Euclidian vacuum. The limit $H=0$ for the ``massless''
conformally coupled scalar field, however, cannot be defined in this
notation \cite{brmo}. In order to obtain the proper behavior of the
``massless'' conformally coupled scalar field in the limit $H=0$, we
must use a system of bounded global coordinates
$(X^\mu,\;\mu=0,1,2,3)$ well suited to describe a compactified
version of dS space, namely S$^3 \times{\rm S}^1$ (Lie sphere)
\cite{garota}. This mode defines the Euclidian vacuum \cite{chta}.
The above procedure, however, cannot be used for the massless vector
field, since the polarization vector which depends on the de Sitter
plane wave, could not be defined in the null curvature limit. It is
important to note that the two-point function of the conformally
coupled scalar field, obtained by the two different methods are one
and the same. Proper choice of vacuum could also be achieved by
imposing the condition that in the null curvature limit, the
two-point function takes the form of its Minkowskian counterpart. By
imposing the following conditions:\begin{itemize}
    \item  setting the vector two-point function to have a maximally
    symmetric form of bi-vectors in the ambient space notation and,
    \item its exact equivalence with the Minkowskian counterpart
    in the null curvature limit,
\end{itemize} the constant vector $Z$ and the
normalization of the vector field are fixed,
\begin{equation}\label{eq:pola1}
Z^{\lambda}\cdot
Z^{\lambda'}=\eta^{\lambda\lambda'},\quad\sum_{\lambda=0}^{3}Z^{\lambda}_{\alpha}Z^{\lambda}_{\beta}
=-\eta_{\alpha\beta}.
\end{equation} $\lambda$ takes four values for different polarizations.
Henceforth the polarization vector can be defined as, \b {\cal
E}(x,\xi,Z,c,\sigma)=
\left(\bar{Z}^{\lambda}+a(c,\sigma)\frac{Z^{\lambda}\cdot x}{x\cdot
\xi}\,\bar{\xi}\right)\equiv{\cal E}^{\lambda}(x,\xi,c,\sigma).\e
This polarization vector satisfies the following relation: \b{\cal
E}^{\lambda}(x,\xi,c,\sigma)\cdot \bar\xi=(a+1)(Z^{\lambda}\cdot x)
(x.\xi) . \e  It can be shown that by the use of Eq.$(4.9)$, the
properties of the dS polarization vector are very similar to the
Minkowskian case:
$$
\sum_{\lambda=0}^{3}{\cal E
}^{\lambda}_{\alpha}(x,\xi,c,\sigma_1){\cal E
}^{\lambda}_{\alpha'}(x',\xi,c,\sigma_2)=$$
\b-\left(\theta_{\alpha}.\theta'_{\alpha'}+
a_2\frac{\theta_\alpha.x'}{x'.\xi} \bar
\xi'_{\alpha'}+a_1\frac{\theta'_{\alpha'}.x}{x.\xi} \bar
\xi_{\alpha}+a_2a_1\frac{x'.x}{(x.\xi )(x'.\xi)} \bar \xi_{\alpha}
\bar \xi'_{\alpha'}\right). \e

Due to the presence of a singularity on the three dimensional
light-like manifold, the dS vector-plane wave solutions ($4.8$) are
not globally defined \cite{brmo}. For a complete determination of
the solutions ($4.8$), one may consider the solutions in the complex
de Sitter space-time $X_H^{(c)}$. The complex de Sitter space-time
is defined
$$ X_H^{(c)}=\{ z=x+iy\in \C^5;\;\;\eta_{\alpha \beta}z^\alpha
z^\beta=(z^0)^2-\vec z.\vec z-(z^4)^2=-1\}$$ \b =\{ (x,y)\in
\R^5\times  \R^5;\;\; x^2-y^2=-1,\; x.y=0\}.\e Let $T^\pm=
\R^5+iV^\pm$ be the forward and backward tubes in $ \C^5$. The
domain $V^+$(resp. $V^-)$
 stems from the causal structure on $X_H$:
\b V^\pm=\{ x\in \R^5;\;\; x^0\stackrel{>}{<} \sqrt {\parallel
\vec x\parallel^2+(x^4)^2} \}.\e Respective intersections with
$X_H^{(c)}$ are
   \b {\cal T}^\pm=T^\pm\cap X_H^{(c)},\e
which will be called forward and backward tubes of the complex dS
space $X_H^{(c)}$. Finally we define the ``tuboid'' on
$X_H^{(c)}\times X_H^{(c)}$ by \b {\cal T}_{12}=\{ (z,z');\;\;
z\in {\cal T}^+,z' \in {\cal T}^- \}. \e Details are given in
\cite{brmo}. When $z$ varies in ${\cal T}^+$ (or ${\cal T}^-$) and
$\xi$ lies on the positive cone ${\cal C}^+$
$$\xi \in {\cal C}^+=\{ \xi \in {\cal C}; \; \xi^0>0 \},$$
the plane wave solutions are globally defined, since the imaginary
part of $(z.\xi)$ has  a fixed sign. The phase is chosen such that
 \b \mbox{boundary value of} \; (z.\xi)^\sigma \mid_{x.\xi>0}>0.\e
Therefore we have \b K_{\sigma,c}^{\xi,\lambda}(z)={\cal
E}^{(\lambda)}(z,\xi,\sigma,c)(z.\xi)^{\sigma},\e in which $z \in
X_H^{(c)} $ and $\xi \in {\cal C}^+$.

\setcounter{equation}{0}
\section{The two-point function}

The two-point function of the massless conformally coupled scalar
field is studied first, in this section. The field operator and the
vacuum states are defined properly to result in this two-point
function. The massless vector two-point function is then calculated.
The vector field operator and the vacuum state are defined to suit
the above two-point function in the next stage. The null curvature
limits of the two-point functions are then discussed. The relations
between the ambient space notation and the intrinsic coordinates are
studied in the final stage.

\subsection{Scalar two-point function}

The Wightman two-point function for a conformally coupled scalar
field is \cite{brmo} $${\cal W}_0(x,x')= c_0 \int_T
d\mu_T(\xi)[(x.\xi)_+^{-1}+e^{i\pi}(x.\xi)_-^{-1}]
[(x'.\xi)_+^{-2}+e^{-2i\pi}(x'.\xi)_-^{-2}]$$ \b =bv \;W_0(z,z')=bv
\;C_0 P_{-1}^{(5)}(z.z'),\e where
$C_0=\frac{\Gamma(2)\Gamma(1)}{2^4\pi^2}=2\pi^2
 c_0$ and \b W_0(z,z')=\frac{1}{8\pi^2}\frac{-1}{1-{\cal Z
}(z,z')}, \;\;{\cal Z }(z,z')=-z.z'.\e  The function
$P_{\sigma}^{(5)}$ is the generalized Legendre function  of the
first kind given by the following integral representation (valid for
$\cos \theta \in \C \setminus  ]-\infty,-1[$) \cite{brmo}: \b
P_{\sigma}^{(5)}(\cos \theta)=\frac{4}{\pi}(\sin
\theta)^{-2}\int_0^\theta \cos[(\sigma+\frac{3}{2})\tau]\sqrt{2(\cos
\tau-\cos \theta)} d\tau.\e This has the interesting property of
$P_{\sigma}^{(5)}=P_{-3-\sigma}^{(5)}$. By determining the boundary
values of the equation $(5.2)$ we obtain \cite{ta,tag} \b {\cal
W}_0(x,x')=\frac{-1}{8\pi^2}\left[\frac{1}{1-{\cal Z}(x,x')}-i\pi
\epsilon(x^0-x'^0)\delta(1-{\cal Z}(x,x'))\right].\e In the theorem
$4.2$ of \cite{brmo}, it has been shown that this two-point function
satisfies the following conditions: a) positivity, b) locality, c)
covariance, and d) normal analyticity . A de Sitter free field can
be defined at this stage.

Using the superposition principle and two solutions of the scalar
field equation, globally defined in the complex de Sitter space,
the general solution of the scalar field is thoroughly defined: \b
\phi(z)=\int_T
 \{\; a(\xi,\sigma_1)(z.\xi)^{\sigma_1}
 +b(\xi,\sigma_2)(z.\xi)^{\sigma_2}\;
\} d\mu_T(\xi),\e where $T$ denotes an orbital basis of ${\cal
C}^+$ and $\sigma_1=-1,\sigma_2=-3-\sigma_1=-2$. $d\mu_T(\xi)$ is
an invariant measure on ${\cal C}^+$ \cite{brmo}. The boundary
value of this equation is the scalar field, defined globally in
the de Sitter space \b \phi(x)=\int_T
 \{\; a(\xi,\sigma_1)[(x.\xi)_+^{\sigma_1}+e^{-i\pi\sigma_1}(x.\xi)_-^{\sigma_1}]
 +b(\xi,\sigma_2)[(x.\xi)_+^{\sigma_2}+e^{i\pi\sigma_2}(x.\xi)_-^{\sigma_2}]\;
\} d\mu_T(\xi),\e where \cite{geshi}$$ (x\cdot
\xi)_+=\left\{\begin{array}{clcr} 0 & \mbox{for} \; x\cdot \xi\leq
0\\ (x\cdot \xi) & \mbox{for} \;x\cdot \xi>0 \\ \end{array}
\right.$$ Since the measure satisfies
$d\mu_T(l\xi)=l^{3}d\mu_T(\xi)$, $a$ and $b$ must satisfy the
homogeneity condition
$$ \;a(l \xi,\sigma)
=l ^{-\sigma-3}a(\xi),\;\;b(l \xi,\sigma) =l
^{-\sigma-3}b(\xi,\sigma).$$ Implementation of these conditions
result in the integral representation ($5.6$) that is independent
of the orbital basis $T$ \cite{brmo}.

As far as irreducible unitary representations of de Sitter group
are concerned, the two solutions of the wave equation (
$(x.\xi)^{\sigma_1}$ and $(x.\xi)^{\sigma_2=-3-\sigma_1})$ are
equivalent to one another. Naturally, solutions are complex
conjugate of each other $(\sigma^*_1= \sigma_2)$ for principal
series representation, since the homogeneity degree of functions
$(\sigma)$ is complex in this case. In the case of complementary
series representation, however, homogeneity degree is real and as
a result the two solutions, in spite of the equivalence of their
corresponding representation $(\sigma_1, \sigma_2=-3-\sigma_1)$,
are not complex conjugate of each other.

Now we define the conformally scalar field operator, which results
in the above two-point function \b \phi(x)=\int_T
 \{\; a(\xi,\sigma_1)[(x.\xi)_+^{\sigma_1}+e^{-i\pi\sigma_1}(x.\xi)_-^{\sigma_1}]
 +a^\dag(\xi,\sigma_2)[(x.\xi)_+^{\sigma_2}+e^{i\pi\sigma_2}(x.\xi)_-^{\sigma_2}]\;
\} d\mu_T(\xi).\e The vacuum state is defined as follows
$$a(\xi,\sigma)|\Omega>=0,$$ which is fixed by imposing the
condition that in the null curvature limit, the Wightman two-point
function, becomes exactly the same as its Minkowskian counterpart.
This vacuum, $|\Omega>$, is equivalent to the Euclidean  vacuum.
``One particle '' states are \b
a^{\dag}(\xi,\sigma)|\Omega>=|\xi,\sigma>.\e The field operator
$(5.7)$ gives the above two-point function $(5.1)$  $$ {\cal
W}_0(x,x')= <\Omega\mid \phi(x)\phi(x')\mid\Omega>. $$  For the
hyperbolic type submanifold, $T_{4}$, the measure is
$d\mu_{T_{4}}(\xi)=d^{3}\vec{\xi}/\xi_{0}$ and the canonical
commutation relations are represented by \begin{equation}
[a(\xi,\sigma),a^{\dagger}(\xi',\sigma')]=\sqrt{c_0}\delta_{\sigma,
-\sigma'-3} \xi^0\delta^3(\vec\xi-\vec\xi').
\end{equation}

\subsection{Vector two-point function}

The general vector two-point function is calculated explicitly at
this stage in the ambient space notation. The vector two-point
function, which is invariant under the conformal group is then
calculated and the Minkowskian limits are discussed. Finally, the
relation between the ambient space notation and the intrinsic
coordinates, are determined.

Similar to the field solution $(4.1)$, the vector two-point
function ${\cal W}_{\alpha \alpha'}(x,x')$ , which is solution of
the wave equation $(2.13)$, can be found simply in terms of scalar
Wightman two-point functions, \b {\cal W}_{\alpha
\alpha'}(x,x')=\langle \Omega,K_{\alpha}(x)K_{\alpha'}(x')\Omega
\rangle=\theta_{\alpha }.\theta'_{\alpha' } {\cal W}_1(x,x')+\bar
\partial_{\alpha} \bar
\partial'_{\alpha'}{\cal W}_2(x,x'),\e where $\bar \partial_{\alpha} \bar
\partial'_{\alpha'}=\bar \partial'_{\alpha'} \bar
\partial_{\alpha}$. The vector two-point function was obtained in terms
of a ``massless'' conformally coupled scalar two-point function
${\cal W}_0(x,x')$ \cite{garota}, $${\cal W}_{\alpha
\alpha'}(x,x')=\theta_{\alpha }.\theta'_{\alpha' } {\cal
W}_0(x,x')-\frac{c}{2(1-c)}\bar
\partial_{\alpha}\left[\bar
\partial . \theta'_{\alpha' }+x.\theta'_{\alpha' }\right]{\cal
W}_0$$ \b +\frac{2-3c}{1-c}\bar \partial_{\alpha}
[Q^{(0)}]^{-1}x.\theta'_{\alpha' }{\cal W}_0\equiv
D_{\alpha\alpha'}(x,x',c){\cal W}_0,\; c\neq 1.\e We can write
this equation in the following form \b {\cal W}_{\alpha
\alpha'}^c={\cal W}_{\alpha
\alpha'}^{\frac{2}{3}}+\frac{\frac{2}{3}-c}{(1-c)}\bar
\partial_{\alpha} [Q^{(0)}]^{-1}\partial.{\cal W}_{\alpha'}^{\frac{2}{3}},\e where
 $$\partial.{\cal W}_{\alpha'}^{\frac{2}{3}}=3
\left(x.\theta'_{\alpha'}+\left[\bar
\partial . \theta'_{\alpha' }+x.\theta'_{\alpha' }\right]\right){\cal
W}_0,\;\; Q^{(0)}\partial.{\cal W}_{\alpha'}^{\frac{2}{3}}=0. $$
These two-point functions can only be defined properly for the
gauge $c=\frac{2}{3}$ since the term $[Q^{(0)}]^{-1}\partial.{\cal
W}_{\alpha'}^{\frac{2}{3}}$ becomes singular.

Now, we consider the case $c=1$, {\it i.e.} the conformally
invariant two-point function, and $c=0$, {\it i.e.} the physical
part of the two-point function. The ``massless'' vector two-point
function, which satisfies the field equation, is obtained as the
boundary value of the analytic two-point function $W_{\alpha
\alpha'}(z,z')$:
\begin{equation}
W_{\alpha \alpha'}(z,z')= c_s\int_T \sum_{\lambda}{\cal
E}^{\lambda}_{\alpha }(z,\xi,c,\sigma_1) \,{\cal
E}^{\lambda}_{\alpha'}(z^{'},\xi,c,\sigma_2)
(z\cdot\xi)^{\sigma_1}(z'\cdot \xi)^{\sigma_2}\,d\mu_T(\xi),
\end{equation} where $\sigma_1+\sigma_2=-3.$
With the  help of Eq.($4.10$), the vector two-point function is
easily expanded in terms of the analytic scalar two-point function
$W_{s}(z,z')$: $$ W_{\alpha \alpha'}(z,z')= c_s\int_T
\sum_{\lambda}\left(\bar{Z}^{\lambda}_\alpha+a_1\frac{Z^{\lambda}\cdot
z}{z\cdot \xi}\,\bar{\xi}_\alpha\right)
\,\left(\bar{Z'}^{\lambda}_{\alpha'}+a_2\frac{Z^{\lambda}\cdot
z'}{z'\cdot \xi}\,\bar{\xi}_{\alpha'}\right)
(z\cdot\xi)^{\sigma_1}( z'\cdot \xi)^{\sigma_2}\,d\mu_T(\xi)
$$  $$ =
\sum_{\lambda}\left(\bar{Z}^{\lambda}_\alpha+\frac{a_1}{\sigma_1}Z^{\lambda}\cdot
z\,\bar{\partial}_\alpha\right)
\,\left(\bar{Z'}^{\lambda}_{\alpha'}+\frac{a_2}{\sigma_2}Z^{\lambda}\cdot
z'\,\bar{\partial'}_{\alpha'}\right)c_s\int_T
(z\cdot\xi)^{\sigma_1}( z'\cdot \xi)^{\sigma_2}\,d\mu_T(\xi)$$ \b
=
\sum_{\lambda}\left(\bar{Z}^{\lambda}_\alpha+\frac{a_1}{\sigma_1}Z^{\lambda}\cdot
z\,\bar{\partial}_\alpha\right)
\,\left(\bar{Z'}^{\lambda}_{\alpha'}+\frac{a_2}{\sigma_2}Z^{\lambda}\cdot
z'\,\bar{\partial'}_{\alpha'}\right)W_{s}(z,z').\e We define the
arbitrary constant tensor $T$ of rank-$2$ as:
$$T_{\beta\gamma}\equiv \sum_{\lambda}Z^{\lambda}_\beta
Z^{\lambda}_\gamma .$$ The equation $(5.14)$ in terms of this
arbitrary tensor can be written in the following form: $$
W_{\alpha \alpha'}(z,z')=T^{\beta\gamma}
\left(\theta_{\alpha\beta}\theta'_{\alpha'\gamma}+\frac{a_1}{\sigma_1}\theta'_{\alpha'\gamma}z_\beta
\,\bar{\partial}_\alpha+\frac{a_2}{\sigma_2}\theta_{\alpha\beta}z'_\gamma\,\bar{\partial'}_{\alpha'}
+\frac{a_1a_2}{\sigma_1\sigma_2}z_\beta z'_\gamma
\bar{\partial}_\alpha\bar{\partial'}_{\alpha'}\right)\times $$ \b
c_s \int_T (z\cdot\xi)^{\sigma_1}( z'\cdot
\xi)^{\sigma_2}\,d\mu_T(\xi)\equiv D_{\alpha
\alpha'}(z,z',c,\sigma) W_{s}(z,z'),\e where $T_{\beta\gamma}$ and
$c_s$ are arbitrary constants and $W_{s}(z,z')$ is the scalar
two-point function.

In the previous paper \cite{gagata2}, we showed that a maximally
symmetric bi-vector in the ambient space notation has the
following form: \b M_{\alpha
\alpha'}(z,z')=\theta_{\alpha}.\theta'_{\alpha'}f({\cal
Z})+(\theta_{\alpha}.z')(\theta'_{\alpha'}.z)g({\cal Z}).\e By
imposing the following conditions:\begin{itemize}
    \item  setting the vector two-point function to have a maximally
    symmetric form of bi-vectors in the ambient space notation and,
    \item its exact equivalence with the Minkowskian counterpart
    in the null curvature limit,
\end{itemize} the constant tensor $T$ and the
normalization constant $c_s$ are fixed:
$$Z^{\lambda}\cdot
Z^{\lambda'}=\eta^{\lambda\lambda'},\;\;\sum_{\lambda=0}^{3}Z^{\lambda}_{\alpha}Z^{\lambda}_{\beta}
=-\eta_{\alpha\beta},\;\;c_{s}=e^{i\pi(\sigma+\frac{3}{2})}\frac{\Gamma(-\sigma)
\Gamma (3+\sigma)}{2^5 \pi^4}, $$ where $\sigma_1\equiv\sigma$. Note
that, the choice of normalization constant corresponds to the
Euclidean vacuum. $W_s(z,z')$ can be written as a hypergeometric
function (see \cite{brmo}):
$$  W_{s}(z,z')= C_{s}\;
_{2}F_{1}\left(-\sigma,3+\sigma;2;\frac{1+\z}{2}\right)= C_{s}
P_{\sigma}^{5}(-\z) \;\quad\mbox{with}\quad C_{s}=\frac{
\Gamma(-\sigma)\; \Gamma (3+\sigma)}{2^4 \pi^2}.$$

The two-point function, as the boundary value of the analytic
two-point functions, can be attained explicitly in terms of the
following class of integral representation $$ {\cal W} _{\alpha
\alpha'}(x,x')= c_s \int_T
d\mu_T(\xi)[(x.\xi)_+^{\sigma}+e^{-i\pi\sigma}(x.\xi)_-^{\sigma}]\times
$$
\begin{equation} [(x'.\xi)_+^{-3-\sigma}+e^{i\pi(\sigma+3)}(x'.\xi)_-^{-3-\sigma}]
\sum_{\lambda =0}^3{\cal E}^{\lambda}_{\alpha }(x,\xi,c,\sigma)
{\cal E}^{\lambda}_{\alpha'} (x',\xi,c,-3-\sigma).\end{equation}
This relation defines the two-point function in terms of global
plane waves on $X_H$. Its explicit form is: $$ {\cal W}_{\alpha
\alpha'}(x,x')=\theta_{\alpha}.\theta'_{\alpha'}\left(1-\frac{a_1a_2}{\sigma_1
\sigma_2 } {\cal Z}\frac{d}{d{\cal Z}}
 \right){\cal W}_s({\cal Z})$$ \b +(\theta_{\alpha}.x')(\theta'_{\alpha'}.x)\left(\frac{a_1a_2}{\sigma_1
\sigma_2 } {\cal Z}\frac{d^2}{d{\cal
Z}^2}-\left(\frac{a_1}{\sigma_1}+\frac{a_2}{\sigma_2}\right)
\frac{d}{d{\cal Z}}
 \right){\cal W}_s({\cal Z}),\e where we have used the identity
$\bar\partial_{\alpha}{\cal W}_s({\cal
Z})=-\left(\theta_{\alpha}\cdot x'\right)\dz {\cal W}_s({\cal
Z})$. For $c\neq 1$,  ${\cal W}_s$ is a conformally coupled scalar
two-point function and for $c=\frac{2}{3}$, the previous result is
obtained \cite{garota}.

The vector two-point functions could be constructed by dS plane
wave functions $(x.\xi)^\sigma$ and $(x.\xi)^{-3-\sigma}$ that are
directly related to irreducible scalar representation of de Sitter
group, {\it i.e.} discrete, complementary and principal series
representations. The two plane waves are equivalent as far as
irreducible representation is concerned. In the case $c=1$, four
sets of solutions are obtained, all related to different values of
$\sigma$ (Eq.$(4.7)$). Two solutions ($\sigma=-1 ,-2$) are related
to the conformally coupled scalar field. The other solutions
($\sigma=-3 ,0$) are related to the minimally coupled scalar
field. In the above formalism, Eq.$(5.18)$, two different vector
two-point functions can be defined, which are not covariant under
the conformal transformation. A conformally covariant vector
two-point function must satisfy equation $(3.19)$. This equation
restricts the solutions of the wave equation corresponding to the
values $\sigma=-2,-3$, which are not equivalent as far as
irreducible representation is concerned. In this case, the
integral representation $(5.15)$ can not be properly defined {\it
i.e.} it depends on the orbits of integration.

To obtain a conformally covariant vector two-point function, the
two-point functions in the form of $(5.10)$ can be utilized to
satisfy the two equations $(3.18)$ and $(3.19)$ simultaneously.
After some algebra, we obtained ${\cal W}_1$, which satisfies:
$$ \frac{d^2}{d{\cal Z}^2}{\cal W}_1=0, \; \mbox{or}\;\;\;{\cal W}_1=C_1+C_2 {\cal
Z}.$$ In order to obtain a regular solution in the large ${\cal
Z}$ domain, we impose the condition $C_2=0$. This results in the
following equation for ${\cal W}_2$:
$$ Q^{(0)}(Q^{(0)}-2) {\cal W}_2=24 C_1{\cal Z},$$ where $$
Q^{(0)}=(1-{\cal Z}^2)\frac{d^2}{d{\cal Z}^2}-4{\cal
Z}\frac{d}{d{\cal Z}}. $$ These solutions are simultaneously
covariant under the conformal group transformation as well as  de
Sitter group. These solutions are associated with a reducible
representation of de Sitter group. By imposing the condition that
the vector two-point function should propagate on the light cone,
we obtained $$ C_1=0,\;\; {\cal W}_2={\cal W}_0,$$ \b {\cal
W}_{\alpha
\alpha'}(x,x')=\left(-\theta_{\alpha}.\theta'_{\alpha'}\frac{d}{d{\cal
Z}}+(\theta_{\alpha}.x')(\theta'_{\alpha'}.x)\frac{d^2}{d{\cal
Z}^2}\right) {\cal W}_0= D_{\alpha\alpha'}{\cal W}_0,\e where
${\cal W}_0$ is a conformally scalar two-point function.

The vector field commutation relation is
$$ iG_{\alpha \alpha'}(x,x')={\cal W}_{\alpha \alpha'}(x,x')-{\cal
W}^*_{\alpha \alpha'}(x,x')$$ \b =[K_\alpha(x),
K_{\alpha'}(x')]=D_{\alpha\alpha'} [\phi(x),
\phi(x')]=D_{\alpha\alpha'}iG(x,x'),\e  where $iG(x,x')$ is the
commutation relation of the conformally coupled scalar field
\cite{brmo,ta,tag}, \b i[\phi(x),\phi(x')]=\frac{H^2}{4¹} \epsilon
(x^0,x'^0) \delta ({\cal Z}(x,x')-1),\e \noindent where ${\cal
Z}(x,x')=-H^2 x.x'=1+\frac{H^2}{2} (x-x')^2  $ and
\b \epsilon (x^0,x'^0)=\left\{\begin{array}{clcr} 1&x^0>x'^0 \\
0&x^0=x'^0\\   -1&x^0<x'^0.\\   \end{array} \right.\e We obtain \b
[K_\alpha(x),
K_{\alpha'}(x')]=\frac{H^2}{4i}D_{\alpha\alpha'}\epsilon (x^0,x'^0)
\delta ({\cal Z}(x,x')-1).\e This field propagates on the light cone
$({\cal Z}=1)$ and the logarithmic singularity dose not appear.
Similar to the scalar field \cite{brmo}, the retarded and advance
propagators are defined respectively by \b G^{ret}_{\alpha
\alpha'}(x,x')=- \theta (x^0-x'^0)G_{\alpha \alpha'}(x,x'),\e \b
G^{adv}_{\alpha \alpha'}(x,x')=\theta(x'^{0}-x^{0})G_{\alpha
\alpha'}(x,x')=G^{ret}_{\alpha \alpha'}(x,x')+ G_{\alpha
\alpha'}(x,x').\e The ``Feynman propagator'' is also defined by $$
iG^{(F)}_{\alpha \alpha'}(x,x')=\langle
\Omega,T\;K_{\alpha}(x)K_{\alpha'}(x')\Omega \rangle$$ \b=
     \theta(x^{0}-x'^{0}){\cal
W}_{\alpha \alpha'}(x,x')+\theta(x'^{0}-x^{0}){\cal W}_{\alpha'
\alpha}(x',x).\e

Using the Wightman two-point function that satisfies the
conditions: a) positivity, b) locality, c) covariance, d) normal
analyticity, e) transversality and d) divergencelessness, we have
already constructed the covariant quantum ``massive'' vector field
in dS space \cite{gata}. In the previous paper \cite{garota}, it
has been shown that for the ``massless'' vector field, we do not
have necessarily the divergenceless condition and as a result we
can not associate this field with a UIR's of the dS group. To
maintain the covariant condition in field quantization we must use
an indecomposable representation of dS group. In this case,
however, we do not have the positivity condition and there appear
unphysical negative and null norm states which are considered  as
the longitudinal ``photons'' and the scalar ``photons''
\cite{garota}. In order to remove the above problems it is
necessary to impose some new conditions similar to Minkowskian
vector field. Following this procedure, positivity and
divergenceless conditions are simultaneously avoided. In contrast
to the ``massive'' vector field it is not evident that two-point
function could be used for construction of ``massless'' quantum
vector field.

However, in order to obtain the above two-point functions, the
vector field operators are defined as:
$$ K_\alpha(x)=\int_T \sum_{\lambda}
 \{\; a(\xi,\sigma_1,\lambda) {\cal
E}_\alpha^{(\lambda)}(x,\xi,c,\sigma_1)[(x.\xi)_+^{\sigma_1}+e^{-i\pi\sigma_1}(x.\xi)_-^{\sigma_1}]
 $$ \b + a^\dag(\xi,\sigma_2,\lambda){\cal
E}_\alpha^{(\lambda)}(x,\xi,c,\sigma_2)[(x.\xi)_+^{\sigma_2}+e^{i\pi\sigma_2}(x.\xi)_-^{\sigma_2}]\;
\} d\mu_T(\xi).\e The vacuum state is defined as follows
$$a(\xi,\sigma,\lambda)|\Omega>=0.$$ This vacuum, $|\Omega>$, is
equivalent to the Euclidean  vacuum. ``One particle '' states are
\b a^{\dag}(\xi,\sigma,\lambda)|\Omega>=|\xi,\sigma,\lambda>.\e
For the hyperbolic type submanifold $T_{4}$ the measure is
$d\mu_{T_{4}}(\xi)=d^{3}\vec{\xi}/\xi_{0}$ and the canonical
commutation relations are
\begin{equation}
[a(\xi,\sigma,\lambda),a^{\dagger}(\xi',\sigma',\lambda')]=-\sqrt{c_s}\delta_{\sigma,
-\sigma'-3}\eta^{\lambda\lambda'} \xi^0\delta^3(\vec\xi-\vec\xi').
\end{equation} It is important to note that the
null curvature limit of this vector field operator is not defined,
however, it dose exist for the vector two-point function.

The Minkowskian limit is now established for the above problem.
First the two-point function of the ``massless'' conformally
coupled scalar field is considered. In contrast to the field
operator, where the null curvature limit ($H=0$) exists only in
the intrinsic notation, the two-point function (Eq.(5.4)), has
Minkowskian limit in both notations (intrinsic notations
\cite{tag} and ambient space notations \cite{ta}). The limit $H=0$
of this equation is \b \lim_{H=0}{\cal W}_0(x,x')={\cal
W}^{(M)}(X,X')=\frac{-1}{8\pi^2}\left[\frac{1}{\mu}+i\pi\epsilon(t-t')\delta(\mu)\right],\;\;
2\mu= (X-X')^2.\e This is exactly the two-point function for
massless scalar field in Minkowski space. For the null curvature
limit the vector two-point function of the Minkowski space is
obtained in the same gauge $c=0$, $$ \lim_{H=0} {\cal W}_{\alpha
\alpha'}(x,x')=\eta_{\mu \nu}{\cal W}^{(M)}(X,X')= {\cal W}_{\mu
\nu}^{(M)}(X,X').$$

Finally, let us write the intrinsic expression of the two-point
functions. In the previous paper \cite{gagata2}, we presented the
following relations between the ambient space and intrinsic
notations
$$Q_{ab'}\equiv\ab {\cal W}_{\alpha\beta'}(x,x'),$$ where
$$
\ab\theta_{\alpha}\cdot\theta'_{\beta'}=g_{ab'}+
(\z-1)n_{a}n_{b'},\qquad \ab {{H^2(\theta'_{\beta'}\cdot
x)\left(\theta_{\alpha}\cdot x'\right)}\over{1-\z^2}}=n_{a}n_{b'}.
$$ $n_{a},n_{b'}$ and $g_{ab'}$ are defined in terms of  the
geodesic distance $\mu(x,x')$, that is the distance along the
geodesic connecting the points $x$ and $x'$. Note that $\mu(x,x')$
can be defined by unique analytic extension even when no geodesic
connects $x$ and $x'$. In this sense, these fundamental tensors
form a complete set. They can be obtained by differentiating the
geodesic distance:
$$n_{a}=\nabla_{a}\mu(x,x'),\qquad n_{a'}=\nabla_{a'}\mu(x,x')$$
and the parallel propagator
$$ g_{ab'}=\sqrt{1-\z^{2}}\nabla_{a} n_{b'}+n_{a}n_{b'}\;.$$
The geodesic distance is implicitly defined \cite{brmo} for
$\z=-H^{2}x\cdot x'$ by
\begin{eqnarray*}
\z&=&\cosh(\mu H)\quad\mbox{for $x$ and $x'$ time-like separated,}
\\
 \z&=&\cos(\mu H)\quad\mbox{for $x$ and $x'$ space-like separated such
 that}\quad \vert x\cdot x'\vert <H^{-2}.
\end{eqnarray*} Therefore, the two-point function in the intrinsic coordinates is
$$ Q_{aa'}=(g_{aa'}+
(\z-1)n_{a}n_{a'})\left(1-\frac{a_1a_2}{\sigma_1 \sigma_2 } {\cal
Z}\frac{d}{d{\cal Z}}
 \right){\cal W}_s({\cal Z})$$ \b +(1-\z^2)n_{a}n_{a'}\left(\frac{a_1a_2}{\sigma_1
\sigma_2 } {\cal Z}\frac{d^2}{d{\cal
Z}^2}-\left(\frac{a_1}{\sigma_1}+\frac{a_2}{\sigma_2}\right)
\frac{d}{d{\cal Z}}
 \right){\cal W}_s({\cal Z}).\e This equation can be written in
 the following form: $$ Q_{aa'}=g_{aa'}\left(1-\frac{a_1a_2}{\sigma_1 \sigma_2 } {\cal
Z}\frac{d}{d{\cal Z}}
 \right){\cal W}_s({\cal Z})$$ \b +n_{a}n_{a'}\left( (\z-1)\left(1-\frac{a_1a_2}{\sigma_1 \sigma_2 } {\cal
Z}\frac{d}{d{\cal Z}}
 \right)+(1-\z^2)\left(\frac{a_1a_2}{\sigma_1
\sigma_2 } {\cal Z}\frac{d^2}{d{\cal
Z}^2}-\left(\frac{a_1}{\sigma_1}+\frac{a_2}{\sigma_2}\right)
\frac{d}{d{\cal Z}}
 \right)\right){\cal W}_s({\cal Z}).\e If we have the two-point function in the intrinsic coordinate  \b
Q_{aa'}(X,X')=g_{aa'}\alpha({\cal Z})+n_{a}\,n_{a'}\beta({\cal
Z}),\e where $\alpha({\cal Z})$ and $\beta({\cal Z})$ are
introduced by Eq.$(4.22)$ in \cite{alja}, the two-point function
in the ambient space notation can be obtained \b {\cal
W}_{\alpha\beta'}(x,x')=\left[\theta_{\alpha}\cdot\theta'_{\beta'}+
{{H^2(\theta'_{\beta'}\cdot x)\left(\theta_{\alpha}\cdot
x'\right)}\over{1+\z}}\right]\alpha({\cal
Z})+\left[{{H^2(\theta'_{\beta'}\cdot x)\left(\theta_{\alpha}\cdot
x'\right)}\over{1-\z^2}}\right]\beta({\cal Z}),\e \b {\cal
W}_{\alpha\beta'}(x,x')=\theta_{\alpha}\cdot\theta'_{\beta'}\alpha({\cal
Z})+H^2(\theta'_{\beta'}\cdot x)(\theta_{\alpha}\cdot x')\left(
\frac{\alpha({\cal Z})}{1+\z}+\frac{\beta({\cal
Z})}{1-\z^2}\right).\e

\setcounter{equation}{0}
\section{Conclusion}

In a series of papers we have shown that the quantization of
various free fields in de Sitter space have a similar pattern of
field quantization as in the Minkowski space. The establishment of
above similarity between the two spaces, is based on the
analyticity in the complexified pseudo-Riemanian manifold. The dS
plane wave solution and the Fourier-Bros transformation in the dS
space play an essential role in the above construction. In this
paper, the conformally invariant wave equations in de Sitter
space, for scalar and vector fields, are introduced and their
solutions and their related two-point functions have been
calculated. We have defined the covariant vector field operator
and the ''particle states'' in the ambient space notation. It is
important to note that although the null curvature limit of this
vector field operator is not defined, the limit of two-point
function dose exist.

The irreducible unitary representations of de Sitter group, which
are associated with rank-2 ``massless'' tensor fields have non
ambiguous extensions to the conformal group $SO(4,2)$. On the
other hand, the irreducible unitary representations of conformal
group are precisely the unique extension of the massless
Poincar\'e group representations with helicity $\pm 2$. In the
quantization process, due to the zero mode problem of the
Laplace-Beltrami operator, de Sitter covariance is broken. To
avoid this problem, a new method was presented for quantization of
the ``massless'' minimally coupled scalar field in dS space-time
\cite{ta,gareta,ta6}. Using this method for linear gravity, the
two-point function is covariant and free of any infrared
divergence \cite{ta2, gagarerota}. In the forthcoming paper, we
shall generalize this construction to the traceless rank-2
``massless'' tensor field (conformal linear quantum gravity in dS
space).

\vskip 0.5 cm \noindent {\bf{Acknowledgement}}: The authors would
like to thank J.P. Gazeau for useful discussion.

\end{document}